# Vascular network remodeling via vessel cooption, regression and growth in tumors

K. Bartha[1] and H. Rieger[2]

[1] *Department of Medical Biochemistry, Semmelweis University, Budapest, Hungary*

[2] *Theoretische Physik, Universität des Saarlandes, 66041 Saarbrücken, Germany*

**Abstract:** The transformation of the regular vasculature in normal tissue into a highly inhomogeneous tumor specific capillary network is described by a theoretical model incorporating tumor growth, vessel cooption, neo-vascularization, vessel collapse and cell death. Compartmentalization of the tumor into several regions differing in vessel density, diameter and in necrosis is observed for a wide range of parameters in agreement with the vessel morphology found in human melanoma. In accord with data for human melanoma the model predicts, that microvascular density (MVD, regarded as an important diagnostic tool in cancer treatment, does not necessarily determine the tempo of tumor progression. Instead it is suggested, that the MVD of the original tissue as well as the metabolic demand of the individual tumor cell plays the major role in the initial stages of tumor growth.

# Introduction

Tumor induced angiogenesis is the formation of new blood vessels around the tumor microenvironment for supporting expansion of the tumor mass (Hanahan and Folkman, 1996). Tumor vessels can arise by angiogenic sprouting (Folkman et al., 1971), intussusception (Folkman et al., 1971), or by recruiting endothelial precursor cells from the bone marrow (Lyden et al., 2001). In addition, tumor cells may also co-opt existing blood vessels from the host (Holash, Maisonpierre et al., 1999).

Tumor growth strictly depends on adequate supply of oxygen ($O_2$) through blood vessels. Vascularization determines pathophysiological characteristics of the tumors, such as tumor invasiveness and metastasis formation. Moreover successful therapy critically depends on the transvascular delivery and permeability of larger molecules into the tumor tissue. Therefore quantification various aspects of the tumor vasculature provides valuable tools for tumor prognosis.

However the degree of vascularization is not a homogeneous measure, it depends on the net balance between proangiogenic and antiangiogenic factors stimulating and inhibiting vessel growth respectively, as well as on nonangiogenic factors, such as $O_2$ and nutrient consumption rates of tumor cells. Recently also the effect of the host microenvironment in the tumor circulation was fully realized (Fukumura et al., 1997). The molecular components involved in tumor induced vascularization include hypoxia-inducible transcription factors, which trigger a coordinated response of angiogenesis by inducing expression of growth factors (GF) (Maxwell et al., 1997; Plate and Risau, 1997). Individual tumors can produce a variety of proangiogenic GFs, among which members of vascular endothelial growth factor (VEGF) and angiopoietin family have a predominant role, all inducing endothelial cell differentiation, proliferation and increase in vessel length [reviewed in: (Carmeliet and Jain, 2000)]. VEGF also exerts morphogenic activity by increasing luminal diameter of existing vessels (Zhou et al., 1998). Angiogenic inhibitors, suppressing proliferation and migration of endothelial cells include thrombospondin-1 (Jiminez et al. 2000), angiostatin (O'Reilly et al., 1994) and endostatin (O'Reilly et al., 1997). They are produced either by the tumor or by the host stromal cells, cause endothelial cell apoptosis and may be involved in tumor dormancy.

Increased cellular proliferation in tumors also results in a switch from $O_2$ consuming to glycolytic pathways providing tumor cells with an alternative energy source during low $O_2$ supply (Iyer et al., 1998; Minchenko et al., 2002). However, each tumor has a minimum $O_2$ consumption rate, characteristic for its metabolic demand, below which tumor cells loose viability, the rate varying with tissue origin and with tumor progression. Thus the minimum $O_2$ and nutrient consumption will limit how far away from the vasculature tumor cells can remain viable, the number of viable tumor cell and their GF production in turn will regulate vascular density, blood flow, and permeability.

Tumor vessels are abnormal in many ways, they show uncontrolled permeability, are tortuous and dilated, have excessive branching, shunts and undergo constant regression and remodelling. This may be due to an imbalance of angiogenic regulators such as VEGF and angiopoietins, giving rise to chaotic blood flow, unstable, leaky vessels (Maisonpierre et al., 1997). The tumor endothelium has widened inter-endothelial junctions, discontinuous or absent basement membrane with loosely associated mural cells (Hashizume et al., 2000). In addition to regulatory mechanisms achieved by angiogenic factors and structural properties of the vessel wall, haemodynamic conditions, especially flow and shear stress (Cullen et al., 2002; Milkiewicz et al., 2001), play a decisive role in maintaining high vascular permeability, inhomogeneous diameter, and concomitant collapse typically observed with tumor blood vessels. It is generally acknowledged, that increase in blood flow contribute to vessel enlargement in any vessel, whereas disturbed flow or severe reduction of blood flow is associated with apoptosis of endothelial cells and vessel regression (Dimmeler and Zeiher, 2000).

Although recent insights in the molecular basis of angiogenesis have resulted in the discovery of many new angiogenic molecules, many questions remain unsolved. So far very little is known about the spatial cues guiding endothelial cells into correct patterns and three-dimensional networks. One of the basic questions is: whether the quantification of some aspects of the vascular network formation can make predictions for tumor progression in patients and if yes, what are these aspects.

From what we have described so far it appears that a theoretical model describing quantitatively the dynamics of the remodeling process of the vascular network during tumor growth should contain at least the following aspects: 1) A pre-existing network representing the normal vasculature that can be modified dynamically via growth, regression and modification of its links (representing blood vessels), 2) A nucleus of tumor cells that can proliferate, age and die, 3) At least two concentration fields: one whose sources are the vessels of the network and represents oxygen and other nutrients, and one whose sources are the tumor cells and which represent GFs, 4) A model for the hemodynamic flow within the network, 5) A set of dynamical rules that describe a) the process of interaction between vessels and tumor cells mediated by the two concentration fields and b) the influence of the blood flow. One intention of our paper is to show that a number of phenomena, like the systematic compartmentalization of the tumor into rapidly vascularizing periphery and necrotic regions and other morphological details of tumor vasculature are already a consequence of the aforementioned five inputs.

Previous attempts to describe mathematically the growth of tumor vasculature either focused on a particular model for angiogenesis in presence of a GF source alone (Anderson and Chaplain, 1998; Levine et al., 2001) or studied tumor growth within a vascular network of fixed topology but with hydrodynamically generated inhomogeneities (Alarcón et al., 2003). From a physicists point of view the tumor/vessel system appears as a dynamically evolving weighted network or graph with a hydrodynamic flow imprinted on it, plus a non-trivial growth process (including birth and death) in continuous or discrete time and space, both interacting non-locally via several diffusion fields. This is a rather complex physical system which has not even been tried to analyze up to now and it turns out to be possible only numerically using methods inspired by computational physics. Therefore we define the model in the next section in terms of the elementary stochastic processes, which in principle could also be formulated in terms of a Master equation (van Kampen, 1992), implement them in a Monte-Carlo simulation and present and discuss the result in the subsequent sections.

## Definition of the model

A clinical case of a human melanoma type tumor was chosen (Döme et al., 2002) as the experimental basis of the theoretical model and other selected experimental data available from solid tumors are also involved. In the model we assume, that tumor induced neovascularization starts with two parallel processes, cooption of existing vessels and vessel sprouting. We do not consider recruiting endothelial cells from the bone marrow.

A hybrid probabilistic cellular automaton model is defined on a square lattice with $N=L^2$ sites with coordinates $\mathbf{r}=(n_x\Delta r, n_y\Delta r)$, $n_x, n_y \in \{1,...,L\}$ and in discrete time $\tau=n\Delta\tau$, with $n=0,1,2,...$ Each site represents an area $\Delta r^2$ with $\Delta r=10\mu m$, which is approximately the size of a single EC or TC, and each time step represents a time interval of length $\Delta\tau=1$hour. Table 1 lists the dynamical variables that define the state of each site at time $\tau$. For instance $e(\mathbf{r},\tau)=1$ (or 0) means that a vessels segment is present (or absent) at site $\mathbf{r}$ and time $\tau$, its radius (if present) is given by $e_r(\mathbf{r},\tau)$, its blood flow rate by $e_Q(\mathbf{r},\tau)$ and its wall shear stress value $e_f(\mathbf{r},\tau)$. If $t(\mathbf{r},\tau)=1$ (or 0) TCs are present (or absent) at site $\mathbf{r}$ and time $\tau$, $t_{uO}(\mathbf{r},\tau)$ is the time it spent in an underoxygenated state. GF and $O_2$ concentration at site $\mathbf{r}$ and time $\tau$ are given by $c_{GF}(\mathbf{r},\tau)$ and $c_{oxy}(\mathbf{r},\tau)$, resespectively. In each time step these variables are modified probabilistically according to the rules listed below, which depend on various fixed parameters listed in table 2.

The initial state of the system at $\tau=0$ represents a regularly vascularized region of a given micro-vascular density $MVD_0$ with a small tumor in the center, c.f. Fig. 1a: Vessels are arranged in a regular pattern, which for simplicity is chosen to be a "Manhattan network" with a lattice constant $a$ (e.g. $a=100\mu m$ yields $MVD_0=100$ vessels per $mm^2$ in 3 dimensions). This means that $e(\mathbf{r},\tau=0)=1$ and $e_r(\mathbf{r},\tau=0)=r_0$ for $\mathbf{r}=(n\cdot a,y)$ and $\mathbf{r}=(x,n\cdot a)$ with $n=0,1,2,...,L/a$ and

x,y=0,1,2,...,L. All other sites have $e(\mathbf{r},\tau=0)=0$. For the initial tumor it is $t(\mathbf{r},\tau=0)=1$ in a region containing $TC_0$ sites determined probabilistically according to the Eden growth rule (Eden, 1961).

In the next time step, and in all subsequent time steps, the following computations are performed one after the other in a Monte Carlo simulation, yielding the new system state at time $\tau+\Delta\tau$, which is evaluated for several quantities of interest. The simulation is stopped either if the tumor reaches the system boundary or if it ceases to exist (depending on the chosen parameter values). We did not incorporate any mechanism resulting in the inhibition or arrest of tumor growth here.

<u>Definition of the GF concentration field $c_{GF}(\mathbf{r},\tau)$:</u> Each TC synthesizes a certain amount of GF and the sum of GF secreted creates a GF concentration field around the tumor. GF (e.g. VEGF) expression is known to be regulated under hypoxic conditions in TC-s (Plate et al., 1992; Shweiki et al., 1992) suggesting that GF concentration might depend on synthesis rate around various TC sites, but the existence of a matrix storage pools for GF (Goerges and Nugent, 2004) would rather predict a generation of a preformed GF gradient in the very close vicinity of TC-s. Therefore in our model we defined GF concentration such a way, that it is a final result of all TC processes and assume a linear decrease of the GF pool of each TC up to a radius $R_{GF}$:

$$c_{GF}(\mathbf{r},\tau) = \Sigma_{\{\mathbf{r'} \text{ with } e(\mathbf{r'},\tau)=1\}} f_{GF}(|\mathbf{r}-\mathbf{r'}|),$$

where the sum is over all TCs at time $\tau$, $f_{GF}(r)=(R_{GF}-r)/N$ for $r \leq R_{GF}$ and $f_{GF}(r)=0$ for $r>R_{GF}$ and the normalization $N$ chosen such that $\Sigma_\mathbf{r} f_{GF}(|\mathbf{r}|)=1$.

<u>Identification of circulated sites $S_{circ}(\tau)$:</u> For each vessel segment (i.e. each site $\mathbf{r}$ with $e(\mathbf{r},\tau)=1$) it is checked whether it is located on an uninterrupted path in the current vessel network that connects the sites (0,0) and (L,L). Technically this check is performed by computing the bi-connected components (Tarjan, 1972) of the graph underlying the vessel network. If this condition is fulfilled, blood can flow through this vessel and $\mathbf{r}$ is a member of the set of circulated vessels $S_{circ}(\tau)$.

Definition of the oxygen concentration field $c_{oxy}(\mathbf{r},\tau)$: Intensive cellular proliferation in tumors results in an increase in oxygen demand, TC-s remain viable within a certain radius around a vessel, generating "cuff"-s on the available oxygen concentration field (Hlatky et al., 2002). However the maximal radius of oxygen concentration field, that still supports TC proliferation ("cuff size") depends both on the oxygen concentration released from a vessel and on the consumption of oxygen by TC-s, modeled by an oxygen concentration threshold $\theta_{oxy}$ defined below. For simplicity we do not calculate $c_{oxy}(\mathbf{r},\tau)$ via the stationary solution of a diffusion equation with vessels as sources and TCs as sinks, but assume that the $O_2$ supply of individual vessel segments depends mainly on circulation.

$$c_{oxy}(\mathbf{r},\tau) = \Sigma_{\{\mathbf{r'} \in S\text{-}circ(\tau)\}} f_{oxy}(|\mathbf{r}\text{-}\mathbf{r'}|),$$

where $f_{oxy}(r)=(R_{oxy}-r)/N'$ for $r \leq R_{oxy}$ and $f_{oxy}=0$ for $r > R_{oxy}$. The normalization constant $N'$ is chosen such that $\Sigma_\mathbf{r} f_{oxy}(|\mathbf{r}|)=1$.

TC proliferation: It is assumed that only TCs with at least one free neighbor site can proliferate. If the local $O_2$ concentration $c_{oxy}(\mathbf{r})$ at such a tumor surface site $\mathbf{r}$ exceeds a threshold $\theta_{oxy}$ this site is occupied with a TC with probability $w(t(\mathbf{r},\tau)=0 \rightarrow t(\mathbf{r},\tau)=1)= \Delta\tau/T_t$ and $c_{GF}(\mathbf{r},\tau)$ is updated. $T_t$ estimates the tumor proliferation time, which can vary enormously with different tumors (Hirst et al., 1982).

EC proliferation, sprouting and sprout migration: Adult normal vessels are quiescent, whereas EC-s in pathologic tumor vessels are stimulated to proliferate and migrate by a local increase in GF concentration to form sprouts (Risau, 1997), migrate and either meet with an other EC forming a tube, or retract (Nehls, 1998). In vascularized tissues like melanoma (Paku, 1998) TC-s also use existing vessels and the high density of anastomosing network (interconnected vascular tubes) arises from an interplay between cooption of old vessels and subsequently induced sprouting (Thompson et al., 1987; Holash, Wiegand, and Yancopoulos, 1999). The latter scenario was applied in our model:

New straight vessel segments between two circulated vessels at site $\mathbf{r}$ and $\mathbf{r}'$ are introduced with probability $\Delta\tau/T_e$ (where $T_e$ is the EC proliferation time) if: $c_{GF}(\mathbf{r},\tau)> \theta_{GF}$, no site in the migration path is occupied by TCs, no site and no neighbor site of the migration path is occupied by ECs except **r** and **r**', and |**r**-**r**'|<$M_{max}$ ($M_{max}$ being the maximum sprout migrgation distance). In case of such an event e(**r**,$\tau$)=1 and $e_r$(**r**,$\tau$)=$r_0$ along this path, and $S_{circ}(\tau)$ and $c_{oxy}$(**r**,$\tau$) are updated.

Vessel dilation: Vessel diameter increases in response to GF, vasodilation is reported to occur concomitantly with capillary growth (Carmeliet and Jain, 2000). Vessel perimeters were found to be increased inside the tumor, in the tumor center EC-s were suggested to participate in vessel dilation (Döme et al., 2002).

In our model a vessel segments at site **r** that is surrounded by TCs and has a GF concentration $c_{GF}$(**r**,$\tau$) larger than $\theta_{GF}$ increase its radius $e_r$(**r**,$\tau$) by an amount $r_0/2\pi$ with probability $\Delta\tau/T_e$ as long as $e_r \leq r_{max}$. To mimic the smoothening effect caused by the surface tension of the vessel walls the location of the dilation is shifted to a neighboring vessel segment if a radius difference larger than $r_0/2\pi$ would arise at the original location.

Blood flow and shear stress computation: The shear stress exerted by the blood flow upon the vessel walls is considered to be a principal stimulus for EC-s, the primary driving force for vessel architecture (Davies, 1995; Ishida et al., 1997). For a given vessel network we identify the vessel segment with cylindrical tubes of radius $e_r$(**r**,$\tau$) and approximate the flow through it by laminar steady Poiseulle flow of a homogeneous liquid. We neglect here the fact that blood is an inhomogeneous fluid which is commonly modeled by a radius dependent viscosity. Then the flow rate is determined by the pressure drop $\Delta P$ between the end points of the segments:

$$e_Q(\mathbf{r},\tau) = Q/\Delta\tau = \mathrm{const} \cdot e_r^4(\mathbf{r},\tau) \cdot \Delta P.$$

The boundary conditions for the pressure are defined in such a way that a homogeneous flow distribution in the vessels of the original network arises: The pressure in the vessel segments decreases linearly on the boundary vessels from $P_{max}$ at **r**=(0,0) to ($P_{max}$-$P_{min}$)/2 at **r**=(L,0) and **r**=(0,L), and from there from ($P_{max}$-$P_{min}$)/2 to $P_{min}$ at **r**=(L,L), resulting in homogeneous global net flow in the diagonal direction. Since only pressure differences enter the flow equations $P_{max}$ and $P_{min}$ can be chosen arbitrarily and are set to 1 and 0, respectively, in arbitrary units. Using Kirchhoff's law we calculate the pressure drop and flow in each vessel segment. The shear stress $e_f(\mathbf{r},\tau)$ acting upon the vessel walls of each vessel is given by $e_f(\mathbf{r},\tau)= \text{const}\cdot e_r(\mathbf{r},\tau) \cdot \Delta P$.

<u>Vessel collapse:</u> Focal necrosis is commonly observed in solid tumors in regions, where the vascular network is inadequate (Griffon-Etienne et al., 1999; Ramanujan et al., 2000). Reduced perfusion of these vessels can be the result of a solid stress of neighboring tumor cells causing the collapse of vessels, or due to apoptosis of ECs induced by local inhibitors of angiogenesis (Dimmeler and Zeiher, 2000). Long-term reduction of wall shear stress is associated with dramatic reduction of the vessel diameter, up to complete vessel occlusion. We used both criteria to identify critical vessels: circulated vessels, which are surrounded by TCs, collapse with probability $p=\Delta\tau/T_{collapse}$ if the wall shear stress $e_f(\mathbf{r},\tau)$ is below a critical value $f_{crit}$, c.f. Fig. 1b-d. After each collapse event $e(\mathbf{r},\tau)\rightarrow 0$ first the set of circulated sites $S_{circ}(\tau)$ and then $c_{oxy}(\mathbf{r},\tau)$ is updated.

<u>EC death:</u> Vessel network sites $\mathbf{r}$ that are not circulated and under-oxygenated ($c_{oxy}(\mathbf{r},\tau)<\theta_{oxy,EC}$) are eliminated with probability ½. $\theta_{oxy,EC}$ is set 10 times higher than the corresponding threshold for TCs, see below.

<u>TC death</u>: TCs adapt their metabolism to hypoxic conditions, therefore $\theta_{oxy,TC}$ is set 10 times lower as $\theta_{oxy,EC}$. TC lifespan under hypoxia and normoxia is difficult to estimate because hypoxia was found to induce apoptosis of TCs but simultaneously lead to selection of p53 deficient colonies induced in hypoxic regions (Yu et al., 2002).

In our model a TC is eliminated with probability ½ only if $O_2$ concentration is under the low (adapted ) $\theta_{oxy,TC}$ for a time $T_{uO}$, c.f. Fig.1c-d. After elimination $c_{GF}(\mathbf{r},\tau)$ is updated.

# Results

**Base case scenario:**

A scenario is chosen to describe a melanoma type tumor, characterized by a small GF threshold ($\theta_{GF}$=0.01), a large GF activation radius ($R_{GF}$=200μm), a large $O_2$ threshold ($\theta_{oxy,TC}$=0.05), a small $O_2$ diffusion radius ($R_{oxy}$=100μm), relatively high vessel collapse probability ($\Delta\tau/T_{collapse}$=0.1 and $f_{crit}$=0.5$f_0$). We set $T_e$ to 40$\Delta\tau$ and $T_t$ to 10$\Delta\tau$. Fig.1. demonstrates the time evolution of the tumor and vessel configuration in the model with the set of parameter values above (see also table 2).

At late stages of the emergence a phenomenological compartmentalization becomes transparent, see Fig. 2: 1) The outer region close to the tumor surface (peritumoral tissue) is highly vascularized by thin vessels, its thickness depends on $\theta_{GF}$ and $R_{GF}$; 2) a well circulated tumor region (tumor periphery) containing vessels grown via sprouting in the outer shell and then enclosed by the growing tumor; 3) the intermediate region inside the tumor with lower MVD and thicker vessels; 4) the necrotic core with only a few large and stable vessels enclosed by a cuff of TCs, the thickness of which depends on $\theta_{oxy}$ and $R_{oxy}$. The border between regions 2 and 3 is diffuse.

A quantitative analysis of this time evolution is performed as follows: At each time MVD, $O_2$ and GF concentration, vessel radius, shear stress, blood flow and pressure inside the vessels are calculated as a function of the radial distance r from the tumor center. Results for the tumor density and MVD is shown in Fig. 3. The peak in the height profile of the tumor density indicates the boundary of the tumor (stochastic fluctuations in the TC proliferation as well as inhomogeneities in the $O_2$ concentration cause the finite width of this step). The tumor radius $R_{tumor}(\tau)$ = max { r | D(r,$\tau$)=1/2 }, also shown in Fig. 3a, grows linearly with time, a fit yields $R_{tumor}(\tau)$ = 0.21·$\tau$ + 60 (time and radius measured in units of $\Delta\tau$ and $\Delta r$, respectively). The radial growth rate (0.21) is approximately twice the TC proliferation rate ($\Delta\tau/T_t$=0.1), which is typical for Eden growth. As long as the peritumoral vessel plexus develops fast enough (small $\theta_{oxy,EC}$, large $R_{GF}$) $R_{tumor}(\tau)$ displays the same time dependence.

At the peak $D_T$ develops a plateau of width ~ $50\Delta r$ before it decreases substantially due to the appearance of necrotic regions in the tumor center. The irregularities in the density profile reflect the exact spatial locations of these necrotic regions, which is random. An average over several realizations of the stochastic process that is described by our model would remove these fluctuations and yield smooth curves. This remark holds for all Figs 3 and 4.

The average growth factor concentration $c_{GF}(r,\tau)$ has approximately the same shape as the tumor density, which is due to the fact that all viable TCs contribute to $c_{GF}(r,\tau)$ in the same way (data not shown).

Fig. 3b shows the MVD, which has a sharp maximum at a radius $R_{MVD}(\tau)$, corresponding to a highly vascularized tumor periphery, which evolves also linearly with time. A fit yields $R_{MVD}(\tau)$ = $0.21\cdot\tau + 60$, implying that it is identical with the tumor radius $R_{tumor}(\tau)$ and demonstrating the appearance of a highly vascularized region in the peritumoral tissue. For radii $r<R_{MVD}(\tau)$ the MVD drops quickly (within ca. $50\Delta r$) to values around the normal tissue $MVD_0$ before slowly decreasing to values significantly lower than $MVD_0$ (corresponding to a poorly vascularized tumor center). The $O_2$ concentration profile $c_{oxy}(r,\tau)$ (not shown) has approximately the same shape as $MVD(r,\tau)$ reflecting the fact that all vessel segments contribute in the same way to $c_{oxy}(r,\tau)$. Hypoxic conditions in the tumor center can thus immediately be read off from the MVD.

The average vessel radius is found to grow linearly with time inside the tumor, as can be seen in Fig. 4a. The maximum vessel radius, set to $r_{max}=3.5\Delta r$, appears as a plateau at sufficiently large times. This behavior is characteristic for a situation in which the GF concentration $c_{GF}(\mathbf{r},\tau)$ inside the tumor ($r<R_{tumor}(\tau)$) is always larger than the actual GF-threshold ($\theta_{GF}$).

The average blood flow per vessel, shown in Fig.4b, increases proportional to $(R_{tumor}(t)-r)^4$ towards the tumor center, due to the fact that the flow depends on the $4^{th}$ power of the vessel radius, which increases linearly according to Fig. 4a. Finer variations of the flow in the peritumoral region are present but not visible on the chosen scale, but see Fig. 4c.

The shear stress F(r,t) acting upon the vessel walls is shown in Fig. 4c. It displays a pronounced dip where the MVD (Fig. 3b) has its maximum, i.e. in the peritumoral plexus. This is due to the fact that an increased MVD in some region reduces the average blood flow per vessel and concomitantly the shear stress F. According to the definition of the model vessels with $F<F_{crit}$ (=$0.5F_0$ here) can collapse resulting in MVD decrease, flow increase and shear stress increase – which is visible in Fig.4c towards the tumor center.

Fig. 4d shows the difference between the blood pressure in the vessel at location (x,y) in the normal vasculature and the blood pressure in the tumor vasculature at time t=600, normalized to the maximum pressure in the upper left corner (x=0,y=0). The blood pressure gradient in the tumor center is up to 50% lower than in normal vessels in spite of the decreased MVD. This is due to dilated vessels with a blood flow capacity up to 200 times larger than that of normal vessels (Fig. 4b).

We determined the fractal dimension $d_f$ of the vascular network with the box-counting method (Mandelbrot, 1982) with boxes ranging from lateral size $1\Delta r$ to $100\Delta r$, the result is shown in Fig. 5, and find it to depend on the tumor region to which the analysis was restricted. The original vasculature has $d_f$=2.0 per construction since the Manhattan pattern represents a compact structure. The complete vasculature that was altered by the tumor has $d_f$=1.85±0.05, concurring with the estimate for the capillary network of various carcinoma (Gazit et al., 1995). The restriction of the analysis to concentric annular rings of width $50\Delta r$ yields a value of $d_f$=1.60±0.05 for the peritumoral region and slightly decreasing values for decreasing radii of the rings reflecting the increasing sparseness of the vessels towards the tumor center. All measurements of fractal aspects of tumor vasculature suffer from the fact that the possible box-sizes span only 2 decades (in real tumors as well as in our model): The minimum size is ca. 10 $\mu$m (approximately the minimum diameter of a capillary) and the maximum is around 1mm (for a tumor of size 5-10mm). Taking into account error bars and transient regimes it is clear that a precise estimate of $d_f$ is hardly feasible. Differences in $d_f$ of the order of 0.1 are, however, significant, and indicate that the local values for $d_f$ correlate strongly with local values for the MVD.

**Parameter dependencies:**

In the following we analyze the various paramter dependencies of the model. Since tumor growth in our model depends strictly on the concentration field $c_{oxy}(\mathbf{r})$ it is usefuld to clarify the shape of this field in the original vasculature, which is shown in Fig. 6 for different values of the lattice constant of the original vasculature *a* (i.e. different values of the original $MVD_0 = 1/a^2$). The oxygen concentration in the original tissue varies spatially only within a few percent when $R_{oxy} \geq a$ and its mean value $c^0_{oxy}$ is proportional to $MVD_0$. TCs at the tumor surface will always survive and proliferate when $\theta_{oxy} < c^0_{oxy}$. On the other hand when $\theta_{oxy} > c^0_{oxy}$ or when $R_{oxy}$ is smaller than *a* the regions between two neighboring vessels in the original network have a $O_2$ content much lower than $c^0_{oxy}$ (or even zero if $R_{oxy} < a/2$), the survival and eventual growth of the tumor becomes probabilistic and depends upon the speed with which new vessels are formed – survival probability decreases drastically in this parameter region and the tumor morphology can become non-circular and even disconnected. In later stages of the tumor growth $R_{oxy}$ and $\theta_{oxy}$ determine the diameter of the cuffs surrounding the large vessels inside the necrotic regions of the tumor.

To clarify the effect of $R_{GF}$ and $\theta_{GF}$ we show the GF concentration field $c_{GF}(\mathbf{r})$ produced by an ideal circular tumor density 1 and radius $50\Delta r$ in Fig. 7a. In the tumor center $c_{GF}(\mathbf{r})=1$ due to the normalization of the function $f_{GF}(r)$. At the boundary the concentration profile decreases in a sigmoidal shape from 1 to 0 over a width of $2R_{GF}$, being 0.5 approximately at the tumor surface. In Fig. 7b we show $c_{GF}$ as a function of the distance r from the tumor surface for different values of $R_{GF}$. $c_{GF}$ scales linearly with $R_{GF}$ which implies that the limit of EC growth, i.e. the width of the region outside the tumor with increased MVD, increases linearly with $R_{GF}$, as is illustrated by the straight line at $c_{GF}=0.05$, from which the maximum width of this region can be read off for the case $\theta_{GF}$.

A non-trivial consequence of large values of $R_{GF}$ (e.g. $40\Delta r$), which produce a wide and dense vasculature in the peritumoral region, is that they are accompanied by larger MVDs also inside the tumor thus inhibiting the emergence of necrotic regions for a large range of collapse probabilities and critical flows - Fig. 8a shows an example of a Monte-Carlo simulation for $R_{GF}=40\Delta r$ and all other parameters as in the base case.

The collapse probability $p_{collapse}=\Delta\tau/T_{collapse}$ and the critical shear stress $f_{crit}$ determine the total necrotic volume and the total vessel number, as is demonstrated by the examples shown in Fig. 8b and Fig. 9. In Fig. 8b all parameters are the same as in Fig. 8a, only $f_{crit}$ is changed from 0.5 to 0.7 , which means that higher flow values are necessary to stabilize vessel. This moderate change decreases MVD inside the tumor drastically and produces large necrotic regions. In Fig. 9 a series of configurations with increasing collapse probability $p_{collapse}$ is shown. The necrotic volume relative to the total tumor mass as well as total number of vessels is shown in Fig. 10 as a function of time for different collapse probabilities ($f_{crit}=0.5$ and the other parameters as in the base case), the necrotic volume relative to the tumor mass starts to increase with time for $t>t_{uO}$ and saturates around $t=500$, ranging from 0.2 for $p_{collapse}=0.05$ to 0.35 for $p_{collapse}=0.5$. This means that the tumor has reached a more or less stationary state in which a constant fraction of the tumor mass is necrotic, except for very small values of $p_{collapse}$, where the necrotic regions are confined to a small region at the tumor center:The data for $p_{collapse}=0.015$ show a slow decrease in the relative necrotic regions for $t>500$. The total number of vessels at time t decreases linearly with time for $p_{collapse}>0.05$, and decreases with increasing $p_{collapse}$. E.g. for $p_{collapse}=0.5$ (and again the other parameters as in the base cae) the total vessel number in the system is 25% less than the total vessel number in the original tissue (in the space region considered) after t=1000 time steps although the MVD is substantially increased in the peritumoral region. Only for small values of $p_{collapse}$ the MVD increases for $t>500$, concomitantly with the decrease in relative necrotic volume: Since vessel collapse is rare the increased MVD in the peritumoral region also survives iinside the tumor.

The TC proliferation time $T_t$ determines the tumor growth speed in well oxidized regions. The EC proliferation time $T_e$ determines the speed with which new vessels are formed in regions with sufficient GF and therefore also effects the tumor growth speed if $\theta_{oxy}$ is smaller than the critical value (0.12 for $MVD_0=100/mm^2$).

## Discussion

Within the base case scenario we have demonstrated that the model reproduces the spatially resolved experimental data of (Döme et al., 2002) for vessel radii and MVD in melanoma. The model can also be adopted to other tumor growth scenarios, ranging from tumors with a low oxygen demand that exclusively coopt pre-existing vessels to tumors that produce a characteristic peritumoral vascular plexus because of high oxygen demand or high GF production rate. It also describes scenarios in which tumors are completely filled with viable TCs and high MVD in the center on one side and morphologies containing large necrotic regions and only a few very thick vessels on the other side.

An important prediction of our model is that large parts of the vascular network can be cut off from the blood circulation by only a few vessel collapses. Consequently the emergence of necrotic regions does not necessarily rely on the presence of large amounts of anti-angiogenic factors (Ramanujan et al., 2000). We find that failure of only a few vital ECs induced by the local mechanical pressure of high tumor densities might also be sufficient to disrupt the oxygen supply of a whole region leading to massive cell death therein. The vessel segments need not only to be connected to the exterior network via an uninterrupted path but at least bi-connected (equivalent to the existence of at least two link-disjoint, uninterrupted paths) in order to be circulated by the blood flow, which lowers the number of collapses that would lead to the necrosis of a region even further.

Vessel collapse events have been shown to correlate with elevated levels of solid pressure exerted by the growing tumor on the intratumoral vessel wall (Boucher and Jain, 1992), which is considered qualitatively in the present version of our model by the criterion that collapse events only occur, if the vessel is surrounded by TCs. Varying stiffness of the surrounded matrix in different tumors can phenomenological be described in our model by varying values for the collapse probability.

In addition to the criterion that vessels are surrounded by TC-s, shear stress was chosen as a second parameter to correlate collapse events. This was motivated by the data obtained on normal vessels, where temporal changes in the haemodynamic flow pattern causing locally a decrease in wall shear stress were shown to lead to a structural reduction of the internal vessel diameter (Pries et al., 1995). We suggest therefore that for a vessel which is surrounded by TC-s (also exerting solid stress) such a contraction could lead to complete collapse.

Another important prediction is that vessel collapse and the local blood flow characteristics have to be correlated via the local shear stress. If the collapse events would occur independent from one another with some probability p, a fundamental law in percolation theory (Stauffer, 1992) predicts that either the interior of the tumor is completely filled (i.e does not contain large connected necrotic regions) or it is completely void up to a small boundary region – except for one special value for the parameter p, the percolation threshold $p_c$. We confirmed this scenario by testing different model variants containing uncorrelated collapse events (see Appendix). The existence of a model parameter like the collapse probability that has to be fine-tuned to a special value in order to reproduce vessels that thread the whole tumor would obviously be unsatisfactory. Only if we correlated the vessel collapse with the shear stress the model predicts a realistic vascular network morphology that is robust against parameter variations. The basic mechanism for this robustness is the redirection of the flow after collapse events into still intact vessels resulting in an increased shear stress in the remaining vessels and thus a drastically reduced collapse probability. Shear stress rather than blood flow as a hemodynamic criterion for vessel stability appears to be plausible, since the ECs in the vessel wall only have (mechanical) information about the shear stress (via the glycocalix) but not on the total flow. We checked that a correlation of vessel collapses with the blood flow also leads to unrealistic network morphologies in which only a few vessels survive within the whole tumor (i.e. a tumor periphery with increased MVD is completely missing). The reason for this is the dependence of the flow from the fourth power of the vessel radius. This leads to a strong variation of the flow between vessels of only slightly different radius implying the survival of only the thickest vessels if collapse is correlated with the flow.

Blood vessels are exposed not only to shear stress but also to the transmural pressure. Since the physiologically relevant difference between the microvasular pressure and interstitial fluid pressure is generally very low in tumor vessels due to their leakiness (Boucher et al., 1996), the pressure-shear hypothesis (Pries et al. 1995) implies an adaptation of the vessel walls towards low equilibrium shear-stress via vasodilation. Together with the effect of GF this mechanism explains the ubiquitous presence of dilated vessels in tumors. Nevertheless blood pressure in tumor vessels remains enigmatic: It is usually larger than the pressure in normal vasculature in spite of the increased MVD and larger vessel diameter in tumors. In addition to geometrical effects also the solid stress exerted by the tumor might play an important role. It appears promising to incorporate solid stress into our model, by using a continuum description for the growing tumor.

The analysis of the parameter dependencies of our model revealed that the morphology of the tumor vasculature depends most strongly on the potential width of the peritumoral plexus (related to the diffusion range of the growth factor molecules) and on the way in which the process of vessel collapse takes place. Varying the probability with which vessels can collapse translates immediately into a variation of the necrotic volume, even more dramatically upon variation of the critical shear force.

A general feature of the remodeling process of the normal vasculature into the tumor vasculature in our model - and we propose this to hold also for in vivo tumors - is that none of the initial characteristics of the original vessel network survives this process: We assumed a original vascular network that consists of capillaries of equal diameter arranged in a regular grid with a given MVD, which guarantees a homogeneous distribution of oxygen and a constant shear stress in all vessels. Once the tumor grows over it, it gets transformed into a compartmentalized network with irregularly arranged dilated vessels and a decreasing MVD from the tumor periphery to the tumor center, resulting in an inhomogeneous oxygen distribution.

Variations in the structural characteristics of the original vascular network will not influence the compartmentalization of the tumor reported here but will certainly have implications for the emerging network morphology and for the absolute values of microvascular pressure. For instance we observe a global directedness of the central vessels in our model, which is a consequence of the boundary conditions for the hydrodynamic pressure that we assumed, which imprints a global net flow in one particular direction through the tumor. The resulting pressure distribution within the whole network has a diagonal gradient, with higher pressures always on the upper left side of the tumor and lower pressure on the lower right side. Natural vascular networks are organized in arterial and venous trees connected by capillary beds, with blood pressure generally decreasing from arteries over arterioles to capillaries and venoules. Their pressure distribution has therefore a structural basis. It would be useful to initialize our model with an original network that is organized in the same way, because then one can also expect to obtain a realistic distribution of the absolute values of the hydrodynamic pressure within tumor vessels. The modulus of the pressure gradient within tumor vessels should, however, remain decreased towards the tumor center, as in the present version of our model.

The fractal analysis of the tumor vasculature predicted by our model showed that for the base case scenario the fractal dimension $d_f$ is around 1.85, which is close to the value found in various carcinoma (Gazit et al., 1995). There it has been emphasized that this value corresponds closely to the one for invasion percolation, which is 1.89 (Stauffer and Aharony, 1992) relating the growth of tumor vasculature to the expansion of a network throughout a medium with randomly distributed heterogeneities in strength. Our model does not involve any heterogeneity aspects of an extracellular matrix surrounding a growing tumor, nevertheless the fractal characteristics of the tumor vasculature turns out to be similar to the one of real tumors.

The tumor radius grows always linear with time in the model and depends strictly on the TC proliferation time. Only in the case of low original MVD it depends on the speed with which the peritumoral region expands into the underoxygenated areas via sprouting. The tumor would stop growing if the MVD in some region of the original tissue is substantially smaller than necessary for TC proliferation, depending on $\theta_{oxy}$ and $R_{oxy}$. In the examples we discussed here it grew linearly because we assumed a homogeneous original tissue. Other possible growth limiting factors include growth inhibitors either produced by the TCs or the host tissue, which could be added as new concentration field into our model, and solid stress generated by a growing tumor in a confined space (Helmlinger et al., 1997) also elevating microvascular pressure (Griffon-Etienne et al., 1999). However involvement of solid stress into the model would lead to a mechanism for growth inhibition only in later stages of tumor development, at present the maximum diameter of the tumor we consider was 5mm.

A correlation between tumor expansion and a local MVD value is visible in our model only in the case when the original MVD is too low to meet the metabolic demand of the TC-s. In this case tumor growth will occur and sustain only if GF production is sufficient to stimulate sprouting in the peritumoral region ($\theta_{GF}$ is small or $R_{GF}$ is large), MVD must be elevated here. In the tumor center the MVD depends on the vessel collapse probability, thus can be larger or smaller than, or equal to the normal tissue MVD. On the other hand if the original MVD is large, there is no correlation between tumor growth and increased (outside) or decreased (inside) levels of MVD, neither locally nor globally. Based on this general behavior of our model we conclude that MVD even when measured in different regions of the tumor is not a reliable diagnostic tool to predict the growth of a single tumor. The correlation found between central MVD of human melanoma and the outcome of the disease (Döme et al., 2002) is probably due to metastasis formation which we did not consider here.

Conclusion: On the basis of a theoretical model, tested on experimental data for human melanoma, it was shown that the microvascular environment of the host is the dominant condition for tumor progression, once this is initiated. Microvascular density within the tumor can be drastically decreased due to the instability of tumor vessels without disturbing the growth at the tumor periphery, implying that MVD measurements might be an unreliable diagnostic tool for tumor progression. The specific reduction of vessel density of the host tissue, in addition to the inhibition of growth factor production of tumor cells, appears to be the most efficient strategy to stop tumor growth.

Acknowledgements: We thank Balázs Döme for his input and discussions. This study was supported by the Hungarian Grant OTKA T-37454.

# Appendix:

Here we consider a variant of our model in which the process of vessels collapse is uncorrelated with the blood flow pattern on the network.

We change the definition of the model introduced in the main text by canceling the flow computation and introduce a fully stochastic vessel collapse with a probability $p_{collpase}$: Circulated vessels, which are surrounded by TCs, collapse with probability $p_{collpase}=\Delta\tau/T_{collapse}$. After each collapse event $e(\mathbf{r},\tau)\to 0$ first the set of circulated sites $S_{circ}(\tau)$ and then $c_{oxy}(\mathbf{r},\tau)$ is updated. Vessels that have a radius that is larger than a certain threshold $r_{stab}$ are supposed to be stable and cannot collapse any more (without any stabilization mechanism all vessels would eventually vanish).

Fig. 11 shows the result of the simulation of this model with for different values for $p_{collpase}$ (and other model parameters as in the base case) and demonstrates that the absence of a correlation between vessel collapse and blood flow leads to unrealistic vessel morphologies and a sharp percolation transition from configurations with small disconnected necrotic regions for small collapse probabilities to configurations consisting only of small annulus of viable tumor cells.

A more detailed analysis (data not shown) reveals a sharp percolation transition takes place at $p_{collapse}=0.049$ which is in the same universality class as conventional percolation (Stauffer and Aharony, 1992): Essentially in this model variant the vessels can collapse only for a fixed time interval – starting at the point when the tumor grows over them and ending at the point when they reach the stabilization radius (note that the radius grows linear in time once vessels are surrounded by TCs). The collapse rate multiplied with this time interval yields the total probability with which a vessel segment is removed. This process leads naturally to a connectivity percolation transition, and much earlier (i.e. for smaller values of $p_{collapse}$) to a bi-connectivity percolation transition.

# References

Alarcón, T., H.M. Byrne, P.K. Maini. 2003. A cellular automaton model for tumour growth in inhomogeneous environment. *J. Theor. Biol.* 225:257-274.

Anderson, A. R. A., M.A.J. Chaplain. 1998. Continuous and discrete mathematical models of tumor-induced angiogenesis. *Bull. Math. Biol.* 60:857-900.

Boucher, Y., R.K. Jain. 1992. Microvascular pressure is the principal driving force for interstitial hypertension in solid tumors - implications for vascular collapse. Cancer Res. 52:5110-5114.

Boucher, Y., M. Leunig, R.K. Jain. 1996. Tumor angiogenesis and interstitial hypertension. Cancer Res. 56:4264-4266.

Burri, P.H., R. Hlushchuk, V. Djonov. 2004. Intussusceptive angiogenesis: Its emergence, its characteristics, and its significance. *Dev. Dyn.* 231:474-488.

Carmeliet, P., R.K. Jain. 2000. Angiogenesis in cancer and other diseases. *Nature* 407:249-257.

Cullen, J. P., S. Sayeed, R.S. Sawai, N.G. Theodorakis, P.A. Cahill, J.V. Sitzmann, E. M. Redmond. 2002. Pulsatile flow-induced angiogenesis - Role of G(i) Subunits. *Arterioscler. Thromb. Vasc. Biol.* 22:1610-1616.

Davies P. F. 1995. Flow-mediated endothelial mechanotransduction. *Physiol. Rev.* 75:519-560.

Dimmeler, S., A.M. Zeiher. 2000. Endothelial cell apoptosis in angiogenesis and vessel regression. *Circ. Res.* 87:434-439.

Döme, B., S. Paku, B. Somlai, J. Tímár. 2002. Vascularization of cutaneous melanoma involves vessel co-option and has clinical significance. *J. Path.* 197:355-362.

Eden, M. (1961) in *Proceedings of the 4$^{th}$ Berkley Symposium Mathematics and Probability*, ed. Neyman, J. (University of California Press, Berkley, 1961), p. 223-240.

Folkman J., M. Bach, J.W. Rowe, F. Davidoff, P. Lambert, C. Hirsch, A. Goldberg, H.H. Hiatt, J. Glass, E. Henshaw. 1971. Tumor angiogenesis – therapeutic implications. *N. Engl. J. Med.* 285: 1182-1186.

Fukumura, D., F. Yuan, W.L. Monsky, Y. Chen, R.K. Jain. (1997). Effect of host microenvironment on the microcirculation of human colon adenocarcinoma. Am.J. Pathol. 151:679-688.

Gazit, Y., D.A. Berk, M. Leunig, L.T. Baxter, R.K. Jain. 1995. Scale-invariant behavior and vascular network formation in normal and tumor tissue. *Phys. Rev. Lett.* 75:2428–2431.

Goerges, A. L., M.A. Nugent. 2004. pH regulates vascular endothelial growth factor binding to fibronectin - A mechanism for control of extracellular matrix storage and release. *J. Biol. Chem.* 279:2307-2315.

Griffon-Etienne, G., Y. Boucher, C. Brekken, H.D. Suit, R.K. Jain. 1999. Taxane-induced apoptosis decompresses blood vessels and lowers interstitial fluid pressure in solid tumors: Clinical implications. Cancer Res. 59:3776-3782.

Hanahan, D., and J. Folkman.1996. Patterns and emerging mechanisms of the angiogenic switch during tumorigenesis. *Cell* 86:353-364.

Hashizume, H., P. Baluk, S. Morikawa, J.W. McLean, G. Thurston, S. Roberge, R.K. Jain, D.M. McDonald. 2000. Openings between defective endothelial cells explain tumor vessel leakiness. Am. J. of Pathol. 156:1363-1380.

Helmlinger, G., P.A. Netti, H.C. Lichtenbeld, R.J. Melder, R.K. Jain. 1997. Solid stress inhibits the growth of multicellular tumor spheroids. *Nat. Biotechnol.* 15 :778-783.

Hirst, D.G., J. Denekamp, B. Hobson. 1982. Proliferation studies of the endothelial and smooth-muscle cells of the mouse mesentery after irradiation. *Cell Tissue Kinet.* 15:251-261.

Hlatky, L., P. Hahnfeldt, J. Folkman. 2002 Clinical application of antiangiogenic therapy: Microvessel density, what it does and doesn't tell us. *J. Nat. Cancer Inst.* 94:883-893.

Holash, J., P. C. Maisonpierre, D. Compton, P. Boland, C.R. Alexander, D. Zagzag, G.D. Yancopoulos, S.J. Wiegand. 1999. Vessel cooption, regression, and growth in tumors mediated by angiopoietins and VEGF. *Science* 284:1994-1998.

Holash, J., S.J. Wiegand, G.D. Yancopoulos. 1999 New model of tumor angiogenesis: dynamic balance between vessel regression and growth mediated by angiopoietins and VEGF. *Oncogene* 18:5356-5362.

Ishida, T., Takahashi, M., Corson, M.A., Berk, B.C. (1997) Fluid shear stress-mediated signal transduction: How do endothelial cells transduce mechanical force into biological responses? *Ann. N. Y. Acad. Sci.* 811, 12-24.

Iyer, N. V., L.E. Kotch, F. Agani, S.W. Leung, E. Laughner, R.H. Wenger, M. Gassmann, J.D. Gearhart, A.M. Lawler, A.Y. Yu, G.L. Semenza. 1998. Cellular and developmental control of O-2 homeostasis by hypoxia-inducible factor 1 alpha. *Genes Dev.* 12:149-162.

Jimenez, B., O. V. Volpert, S. E. Crawford, M. Febbraio, R. L. Silverstein, N. Bouck. 2000. Signals leading to apoptosis-dependent inhibition of neovascularization by thrombospondin-1. *Nat. Med.* 6:41-48.

Levine, H. A., S. Pamuk, B.D. Sleeman, M. Nilsen-Hamilton. 2001. Mathematical modeling of capillary formation and development in tumor angiogenesis: Penetration into the stroma. *Bull. Math. Biol.* 63:801-863.

Lyden, D., K. Hattori, S. Dias, C. Costa, P. Blaikie, L. Butros, A. Chadburn, B. Heissig, W. Marks, L. Witte, Y. Wu, D. Hicklin, Z.P. Zhu, N.R. Hackett, R.G. Crystal, M.A.S. Moore, K.A. Hajjar, K. Manova, R. Benezra, S. Rafii. 2001. Impaired recruitment of bone-marrow-derived endothelial and hematopoietic precursor cells blocks tumor angiogenesis and growth. *Nat. Med.* 7: 1194-1201.

Mandelbrot, B. B. 1983. *The Fractal Geometry of Nature*: W. H. Freeman, New York.

Maisonpierre, P.C., C. Suri, P.F. Jones, S. Bartunkova, S. Wiegand, C. Radziejewski, D. Compton, J. McClain, T.H. Aldrich, N. Papadopoulos, T.J. Daly, S. Davis, T.N. Sato, G.D. Yancopoulos. 1997. Angiopoietin-2, a natural antagonist for Tie2 that disrupts in vivo angiogenesis. *Science* 277:55-60.

Maxwell, P.H., G.U. Dachs, J.M. Gleadle, G.L. Nicholls, A.L. Harris, I.J. Stratford, O. Hankinson, C.W. Pugh, P.J. Ratcliffe. 1997. Hypoxia-inducible factor-1 modulates gene expression in solid tumors and influences both angiogenesis and tumor growth. *Proc. Natl. Acad. Sci. USA* 94:8104-8109.

Milkiewicz, M., M.D. Brown, S. Egginton, O. Hudlicka. 2001 Association between shear stress, angiogenesis, and VEGF in skeletal muscles in vivo. *Microcirculation* 8, 229-241.

Minchenko, A., I. Leshchinsky, I. Opentanova, N.L. Sang, V. Srinivas, V. Armstead, J. Caro. 2002. Hypoxia-inducible factor-1-mediated expression of the 6-phosphofructo-2-kinase/fructose-2,6-bisphosphatase-3 (PFKFB3) gene - Its possible role in the Warburg effect. J. Biol. Chem. 277:6183-6187.

Nehls, V., R. Herrmann, M. Huhnken. 1998 Guided migration as a novel mechanism of capillary network remodeling is regulated by basic fibroblast growth. *Histochem. Cell Biol.* 109:319-329.

O'Reilly, M.S., L. Holmgren, Y. Shing, C. Chen, R.A. Rosenthal, M. Moses, W.S. Lane, Y.H. Cao, E.H. Sage, J. Folkman.1994. Angiostatin – a novel angiogenesis inhibitor that mediates the suppression of metastatses by a Lewis lung-carcinoma. *Cell* 79:315-328.

O'Reilly, M.S., T. Boehm, Y. Shing, N. Fukai, G. Vasios, W.S. Lane, E. Flynn, J.R. Birkhead, B.R. Olsen, J. Folkman. 1997. Endostatin: An endogenous inhibitor of angiogenesis and tumor growth. *Cell* 88:277-285.

Paku, S. 1998. Current concepts of tumour-induced angiogenesis. *Pathol. Oncol. Res.* 4:62-75.

Plate, K. H., G. Breier, H.A. Weich, W. Risau. 1992. Vascular endothelial growth-factor is a potential tumor angiogenesis factor in human gliomas invivo. *Nature* 359:845-848.

Plate, K. H., and W. Risau.1995. Angiogenesis in malignant gliomas. *Glia* 15:339-347.

Pries, A.R., T.W. Secomb, P. Gaehtgens. 1995. Design principles of vascular beds. Circ. Res. 77:1017-1022.

Ramanujan, S., G.C. Koenig, T.P. Padera, B.R. Stoll, R.K. Jain. 2000. Local imbalance of proangiogenic and antiangiogenic factors: A potential mechanism of focal necrosis and dormancy in tumors.  Cancer Res. 60:1442-1448.

Risau, W. 1997. Mechanisms of angiogenesis. *Nature* 386:671-674.

Shweiki, D., A. Itin, D. Soffer, E. Keshet. 1992. Vascular endothelial growth-factor induced by hypoxia may mediate hypoxia-initiated angiogenesis. *Nature* 359:843-845.

Stauffer, D. and A. Aharony. 1992. Introduction to Percolation Theory*,* 2nd ed. Taylor & Francis, London.

Tarjan R. 1972. Depth-first search and linear graph algorithms. SIAM J. Comput. 1:146-160.

Thompson, W.D., K.J. Shiach, R.A. Fraser, L.C. Mcintosh, J.G. Simpson. 1987. Tumors acquire their vasculature by vessel incorporation, not vessel ingrowth. *J. Path.* 151:323-332.

Van Kampen, N.G. 1992. Stochastic processes in physics and chemistry. North Holland, Amsterdam

Yu, J. L., J.W. Rak, B.L. Coomber, D.J. Hicklin, R.S. Kerbel. 2002. Effect of p53 status on tumor response to antiangiogenic therapy. Science 295:1526-1528.

Zhou, M., R.L. Sutliff, R.J. Paul, J.N. Lorenz, J.B. Hoying, C.C. Haudenschild, M.Y. Yin, J.D. Coffin, L. Kong, E.G. Kranias, W.S. Luo, G.P. Boivin, J. Duffy, S.A. Pawlowski, T. Doetschman. 1998. Fibroblast growth factor 2 control of vascular tone. *Nat. Med.* 4:201-207.

# Tables

**Table 1:** Variables defining the state at site **r** at time $\tau$

| e | 0/1: EC or vessel absent/present |
|---|---|
| $e_r$ | $\in [0, r_{max}]$, radius of vessel |
| $e_Q$ | $\geq 0$, blood flow rate through vessel segment |
| $e_f$ | $\geq 0$, shear stress on vessel wall |
| t | 0/1 : TC absent/present |
| $t_{uO}$ | $\in [0, T_{max}]$, time of TC in underoxygenated state |
| $c_{oxy}$ | $\geq 0$; oxygen concentration field |
| $c_{GF}$ | $\geq 0$; growth factor concentration field |

**Table 2:** Model parameters

| $\Delta r$ | lattice constant: space discretization | 10 μm |
|---|---|---|
| $\Delta \tau$ | length of time step: time discretization | 1h |
| $MVD_0$ | original MVD | $100/mm^2$ |
| $TC_0$ | original tumor size | $10^3$-$10^5$ |
| $R_{GF}$ | growth factor diffusion radius | ~100-400 μm |
| $R_{oxy}$ | oxygen diffusion radius | ~100-150 μm |
| $\theta_{GF}$ | growth factor threshold | $10^{-3}$-$10^{-1}$ |
| $\theta_{oxy,EC}$ | oxygen threshold for Ecs | $10^{-2}$-$10^{-1}$ |
| $\theta_{oxy,TC}$ | oxygen threshold for TCs | $0.1 \cdot \theta_{oxy,EC}$ |
| $T_t$ | TC proliferation time | $10\Delta\tau=10h$ |
| $T_e$ | EC proliferation time | $40\Delta\tau=40h$ |
| $r_0$ | initial vessel radius | $1\Delta r=10$ μm |
| $r_{max}$ | maximum vessel radius | $3.5 r_0=35$ μm |
| $M_{max}$ | maximum sprout migration distance | $10\Delta r=100$ μm |
| $T_{collapse}$ | collapse time of critical vessels | $10-100 T_e$ |
| $f_{crit}$ | critical shear stress on vessel walls | $0.2-0.8\ f_0$ |
| $T_{uO}$ | maximum TC survival time in hypoxia | $100\Delta\tau=100h$ |

# Figure legends

**Fig.1a-d:** Series of images showing the time evolution of the tumor and vessel configurations in our model at different time steps.

The normal vasculature, to be seen far away from the tumor, is characterized by straight lines arranged in a grid like network with a characteristic line-to-line distance (here 100μm) defining the normal MVD (here 100/mm$^2$). The color of the blood vessels indicate their blood flow. Blue corresponds to normal, red to high and yellow to low flow values. The thickness of the lines represents the vessel radius. Sites occupied by TCs are color coded in green to black, increasing darkness indicating their age. White regions are empty sites, i.e. inside the tumor the necrotic regions. For the parameter values see the text and table 2. Only the part of the whole system (L=512) containing the peritumoral region is shown.

**(a)** One time step (t=1) after the initialization of the system: The irregular structure of the initial tumor is typical for an Eden cluster (Eden, 1961). One sees that 5 new vessels are formed within one time step at the boundary. **(b)** t=50: Extensive formation of new vessels can be seen at the peritumoral region. New vessels frequently have lower blood flow values than vessels in the original network. The original vessels are almost all intact. **(c)** t=100: The progressive tumor growth pushes the peritumoral plexus further into the normal tissue. Inside the tumor, MVD is drastically reduced leaving large regions of the tumor underoxygenated. **(d)** t=200: TCs inside the low MVD region visible in (c) died after being 100 times steps in an underoxygenatd state leaving a large necrotic region to be seen as a white spots.

**Fig. 2:** The state of the system with the same parameters as in Fig. 1 after t=1000 time steps. The color coding is the same as in Fig. 1. The lateral size of the region shown is 5.1mm.

**Fig. 3:** As a function of the radial distance r from the tumor center (measured in units of Δr) and the time t (measured in units of Δt) it is shown: **(a)** the average tumor density $D_T(r,t)$ as defined by the average number of TCs per lattice sites at time t, **(b)** average microvascular density MVD(r,t), the average number of vessels per lattice site, relative to normal tissue $MVD_0$. The average is done over an annulus of width 10Δr with central radius r.

**Fig. 4:** As in Fig. 3 it is shown as a function of the radial distance r from the tumor center and the time t: **(a)** the average vessel radius R(r,t), **(b)** the blood flow per vessel $Q(r,t)=e_q(r,t)$, **(c)** the shear stress $F(r,t)=e_f(r,t)$. **(d)** The difference in the blood pressure inside the vessels (with respect to normal). Since the local blood pressure inside the vessels does not have the radial symmetry of the other quantities studied in a-c we show the pressure field P(x,y) at time t=600 minus the normal pressure $P_0$(x,y) at t=0 (i.e. for normal vascularization), where (x,y) are the coordinates of the vessels. One recognizes the spatial location and extent of the circular tumor in the middle. The global flow direction, enforced via the boundary conditions, is from left (0,0) to right (512,512). Looking along this direction, the pressure is decreased in front of the tumor (left) and increased at the back of the tumor (right).

**Fig. 5:** Determination of the fractal dimension $d_f$ of the vessel network at time t in the base case (Fig. 2 ) via the box-counting method: The number of Boxes of size L that is needed to cover completely the vasculature is plotted as a function of L in log-log scale. The slope $d_f$ of the curve is the fractal dimension. We confined the measurement to annuli with fixed outer radius that is determined by the limit of the peritumoral plexus and with varying inner radius $R_i$. The slope of the curves decreases with increasing $R_i$: $d_f$=1.85+/-0.05 for $R_i$ =0 (full squares, which corresponds to the complete tumor vasculature) and $d_f$=1.60+/-0.05 for $R_i$ =200 (which corresponds to the peritumoral plexus exclusively), indicating that the fractal dimension is not a homogeneous measure over all regions of the tumor vasculature.

**Fig. 6:** Oxygen concentration field $c_{oxy}$(x,y) of the original vasculature for different lattice constants *a* (i.e. different $MVD_0$) and different values for $R_{oxy}$, which is chosen to be equal to *a*. The variation of $c_{oxy}$ between the vessels is only a few percent and the average value decreases with decreasing a (or $MVD_0$).

**Fig. 7: a)** (Top) Growth factor concentration field $c_{GF}$(x,y) for an ideal circular tumor of density one and radius $50\Delta r$ and $R_{GF}$=$20\Delta r$. **b)** (Bottom) $c_{GF}$ as a function of the distance from the tumor surface for the ideal tumor of a for different $R_{GF}$.

**Fig. 8:** Effect of a large value of RGF on the tumor morphology. **Top:** shows a configuration after 1000 time steps for RGF=$40\Delta r$ but all other parameters as in the base case. Compared to Fig. 2 the necrotic regions are much smaller and the MVD is homogeneously increased with the tumor. (Bottom): Parameters as in the top figure except the critical shear stress $f_{crit}$, which is now 0.7 (instead of 0.5). Whereas the morphology in the top figure is rather stable even for large collapse probabilities only a small variation of $f_{crit}$ produces large necrotic regions and decreased MVD within the tumor.

**Fig. 9:** Sequence of configurations with increasing collapse probability (from left to right it is $p_{collapse}$=0.02, 0.04, 0.1, 0.5) and $f_{crit}$=0.5. Here we have chosen $R_{GF}$=10, $\theta_{GF}$=0.01 and the other parameters as in the base case. The same seed for the random number generator for the simulations in all cases is chosen to emphasize the effect of the variation of $p_{collapse}$.

**Fig.10: a)** The total volume of the necrotic regions with the tumor divided by the total tumor mass as a function of time for different collapse probabilities (parameters as in Fig. 9). **b)** The relative deficit/excess volume of the vasculature as a function of time for different collapse probabilities (parameters as in Fig. 9).

**Fig. 11:** Tumor and vessel configurations after time t=2000 for the model with uncorrelated vessel collapse (i.e. without blood flow and without a critical shear foce determining the vessel collpase) – from left to right it is $p_{collapse}$=0.04, 0.05 and 0.06. The morphology of the vessel network is much more tortuous, in particular a global directedness of the vessel is absent and only for values of $p_{collapse}$ around 0.05 the inner region of the tumor is neither completely filled nor completely void with thick vessels threading the tumor.

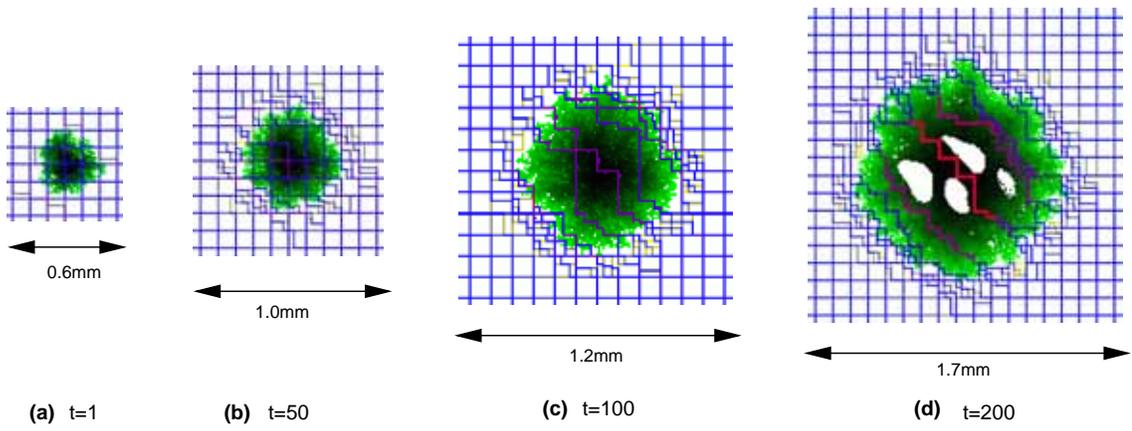

**(a)** t=1   **(b)** t=50   **(c)** t=100   **(d)** t=200

Fig. 1

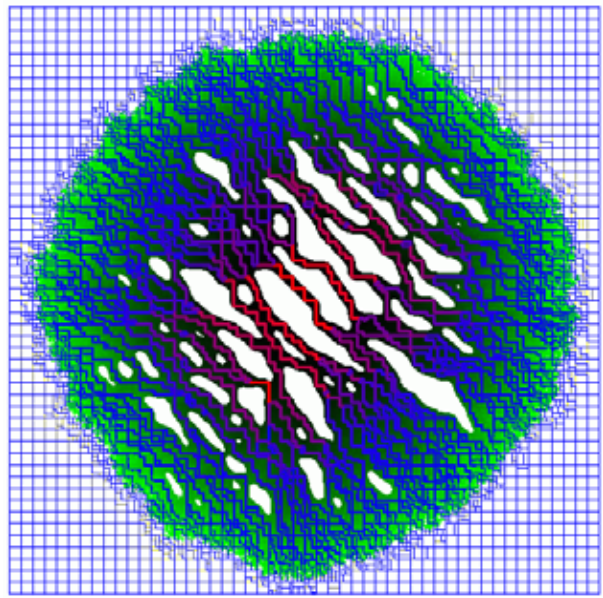

Fig. 2

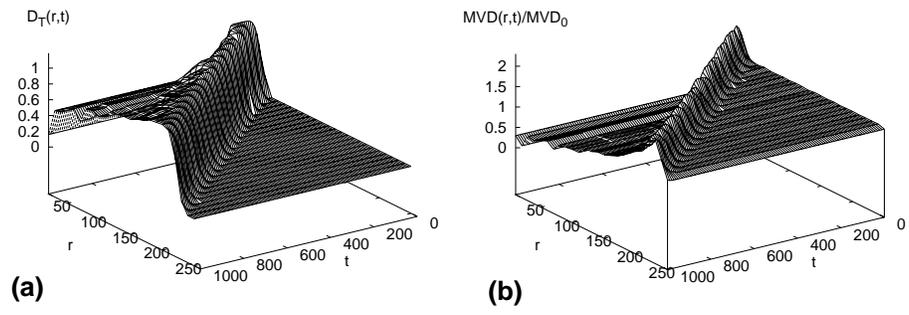

Fig. 3

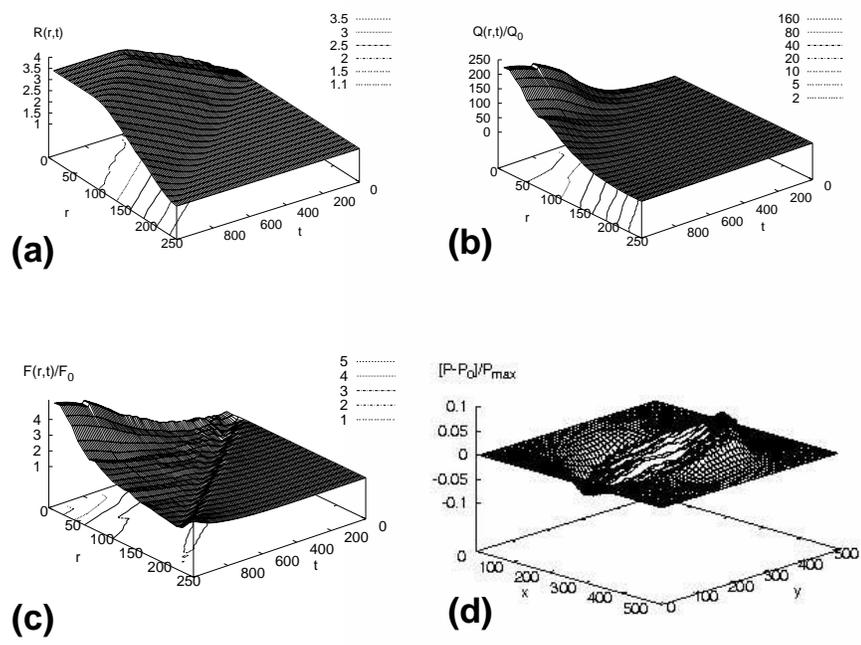

Fig. 4

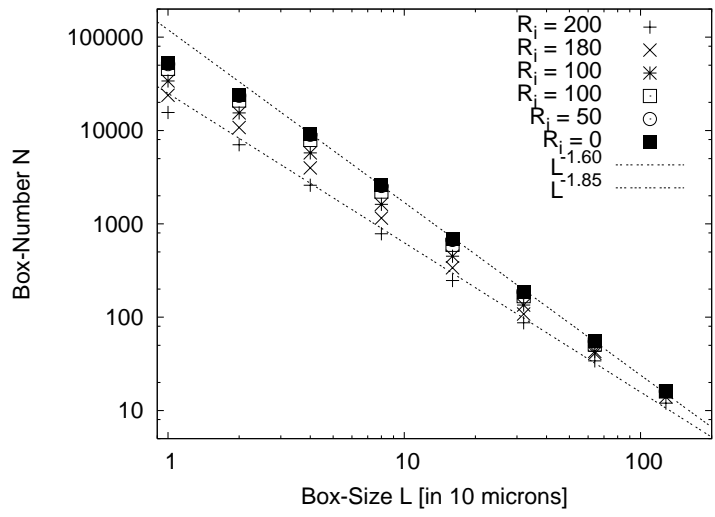

**Fig. 5**

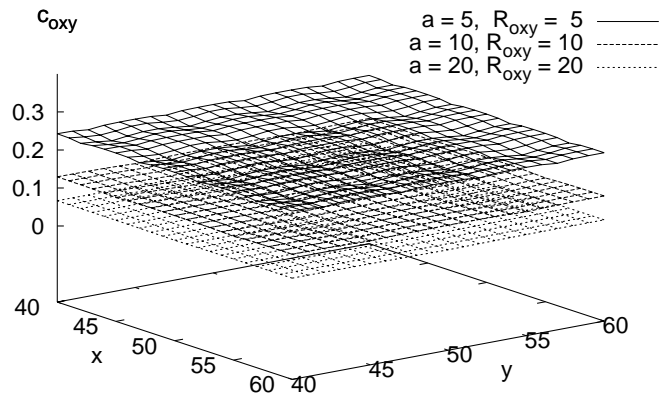

**Fig. 6**

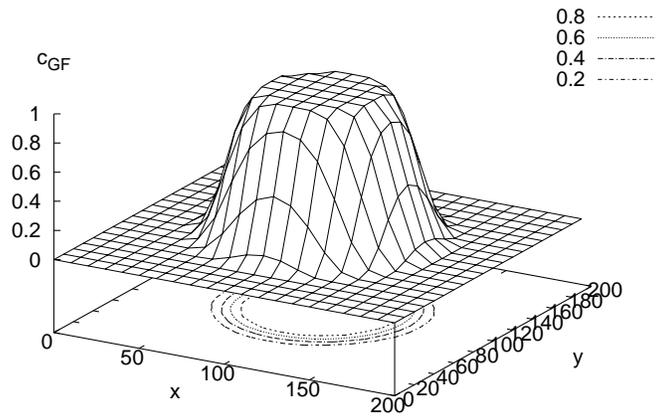

**Fig. 7a**

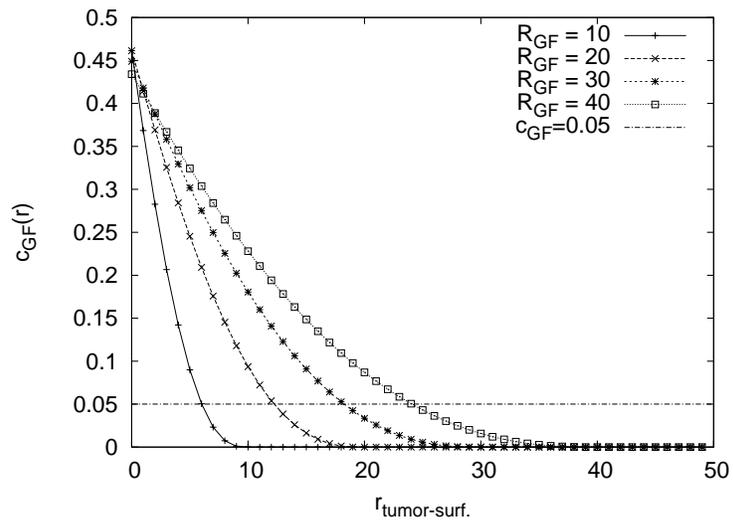

**Fig. 7b**

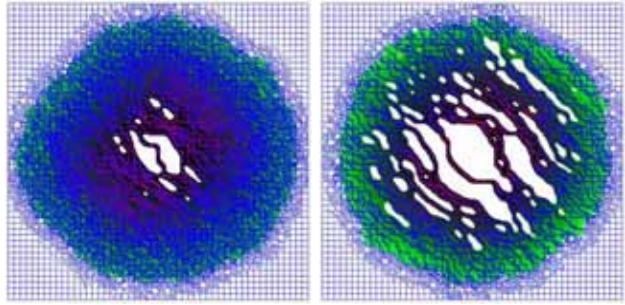

Fig. 8

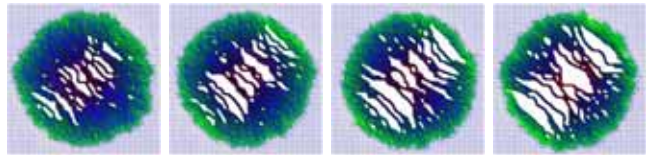

Fig. 9

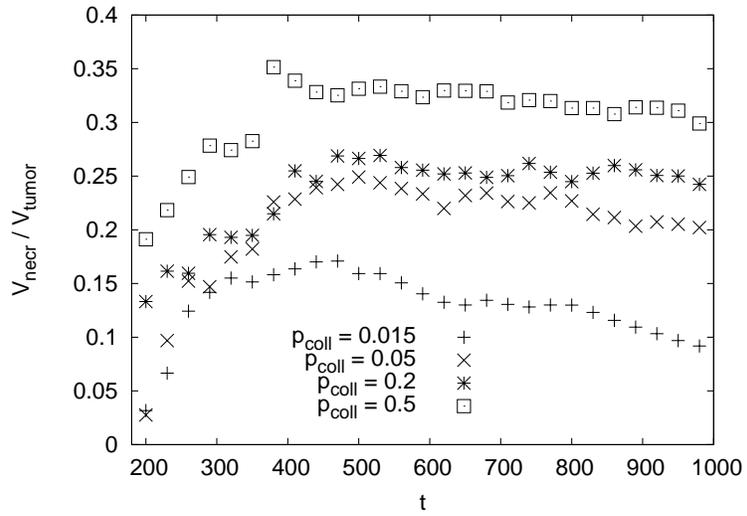

**Fig. 10a**

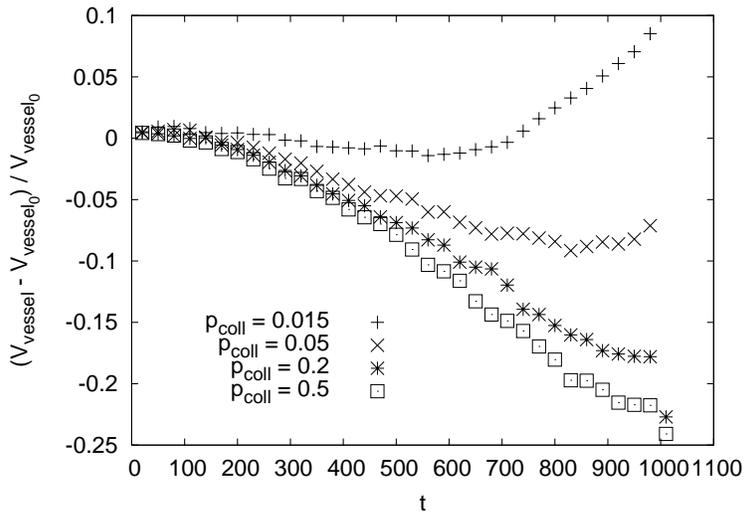

**Fig. 10b**

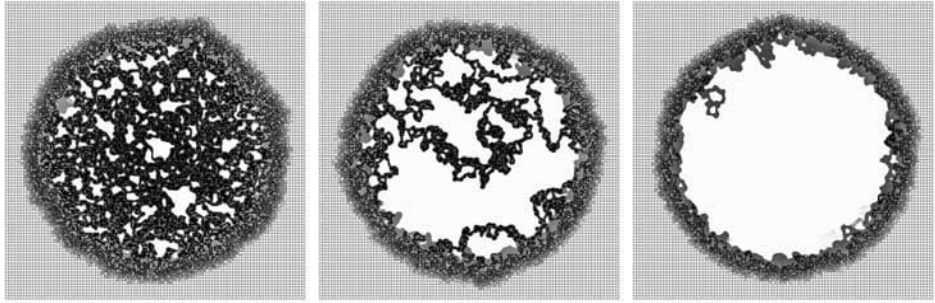

Fig. 11